\documentclass[prd,amsmath,amssymb,showpacs,superscriptaddress,nofootinbib,twocolumn]{revtex4-1}
\usepackage{graphicx}
\usepackage{epsfig,graphics,subfigure,psfrag,amsmath,amssymb}
\usepackage{lineno}
\usepackage{dcolumn}
\usepackage{bm}
\usepackage{overpic}
\usepackage{xspace}

\usepackage{rotating}
\usepackage{color}
\usepackage{multirow}
\usepackage{colortbl}
\usepackage{epstopdf}
\usepackage[colorlinks,linkcolor=blue,anchorcolor=blue,citecolor=blue]{hyperref}

\newcommand{\BESIII}{BES\uppercase\expandafter{\romannumeral3}\xspace}

\begin{document}
%\title{\boldmath Study of $\psi(3686)\to\ppphi$}
\title{\boldmath Observation of $\eta^\prime\rightarrow\pi^+\pi^-\mu^+\mu^-$}
%===========================================================================
%============================ Authors ======================================
%===========================================================================
\author{M.~Ablikim$^{1}$, M.~N.~Achasov$^{10,c}$, P.~Adlarson$^{64}$, S. ~Ahmed$^{15}$, M.~Albrecht$^{4}$, A.~Amoroso$^{63A,63C}$, Q.~An$^{60,47}$, X.~H.~Bai$^{54}$, Y.~Bai$^{46}$, O.~Bakina$^{29}$
% ...
\\
\vspace{0.2cm}
(BESIII Collaboration)\\
\vspace{0.2cm} {\it
$^{1}$ Institute of High Energy Physics, Beijing 100049, People's Republic of China\\
% ...
$^{4}$ Bochum Ruhr-University, D-44780 Bochum, Germany\\
% ...
$^{10}$ G.I. Budker Institute of Nuclear Physics SB RAS (BINP), Novosibirsk 630090, Russia\\
% ...
$^{15}$ Helmholtz Institute Mainz, Johann-Joachim-Becher-Weg 45, D-55099 Mainz, Germany\\
% ...
$^{29}$ Joint Institute for Nuclear Research, 141980 Dubna, Moscow region, Russia\\
% ...
$^{46}$ Southeast University, Nanjing 211100, People's Republic of China\\
$^{47}$ State Key Laboratory of Particle Detection and Electronics, Beijing 100049, Hefei 230026, People's Republic of China\\
% ...
$^{54}$ University of Jinan, Jinan 250022, People's Republic of China\\
% ...
$^{60}$ University of Science and Technology of China, Hefei 230026, People's Republic of China\\
% ...
$^{63}$ University of Turin and INFN, (A)University of Turin, I-10125, Turin, Italy; (B)University of Eastern Piedmont, I-15121, Alessandria, Italy; (C)INFN, I-10125, Turin, Italy\\
$^{64}$ Uppsala University, Box 516, SE-75120 Uppsala, Sweden\\
% ...
$^{c}$ Also at the Novosibirsk State University, Novosibirsk, 630090, Russia\\
% ...
}
}
%End authors

\date{\today}

\begin{abstract}
  Using $(1310.6 \pm 7.0) \times 10^{6}~J/\psi$ events acquired with the BESIII detector at the BEPCII storage rings, the decay $\eta^\prime\rightarrow\pi^+\pi^-\mu^+\mu^-$ is observed for the first time with a significance of 8$\sigma$ via the process $J/\psi\rightarrow\gamma\eta'$. We measure the branching fraction of $\eta^\prime\rightarrow\pi^+\pi^-\mu^+\mu^-$ to be $\mathcal{B}(\eta^\prime\rightarrow\pi^+\pi^-\mu^+\mu^-)$=(1.97$\pm$0.33(stat.)$\pm$0.18(syst.))$\times10^{-5}$, where the first and second uncertainties are statistical and  systematic, respectively. 
\end{abstract}

\maketitle
\section{INTRODUCTION}

The decays $\eta^\prime\rightarrow \pi^+\pi^-\ell^+\ell^-$ (with $\ell$ = $e$ or $\mu$), which are expected to proceed via a virtual photon intermediate state, are especially interesting since these two decays may involve the box anomaly contribution~\cite{intro03} and could be used to test the possibility of double vector meson dominance. 
Theoretically these decays have been investigated with different models, including the effective meson theory~\cite{intro05}, the chiral unitary approach~\cite{intro06} and the hidden gauge model~\cite{intro07}. Due to the larger muon mass, the virtual photon conversion to dimuon is significantly suppressed relative to the conversion to dielectron. 
Therefore, the predictions for the branching fraction of $\eta^\prime\rightarrow\pi^+\pi^-\mu^+\mu^-$ are in the range of $(1.5-2.5) \times 10^{-5}$~\cite{intro05, intro06, intro07}, which are about two orders of magnitude lower than those for $\eta^\prime\rightarrow\pi^+\pi^-e^+e^-$. This explains why only $\eta^\prime\rightarrow\pi^+\pi^-e^+e^-$, with a branching fraction of $(2.11\pm0.12)\times10^{-3}$~\cite{intro09}, has been observed to date. 

In previous analyses, the CLEO collaboration~\cite{intro08} used $4\times10^4$ $\eta'$ from the decay chain $\psi(2S)\rightarrow\pi^+\pi^-J/\psi, J/\psi\rightarrow\gamma\eta'$, while the BESIII analysis~\cite{intro09} used $1.2\times10^6$ $\eta'$ from $J/\psi\rightarrow\gamma\eta'$. Both CLEO and BESIII have performed searches for $\eta^\prime\rightarrow\pi^+\pi^-\mu^+\mu^-$~\cite{intro08,intro09}, but no significant signal was observed. The most stringent upper limit of $\mathcal{B}(\eta^\prime\rightarrow\pi^+\pi^-\mu^+\mu^-)<2.9\times 10^{-5}$ at the 90\% confidence level, is provided by the BESIII experiment. This upper limit lies in the same order of magnitude as the theoretical predictions. In this paper we analyse the sample of $1.31 \times 10^9$ $J/\psi$ events~\cite{sys01}, which is about five times larger than the subsample used in the previous BESIII measurement and enables us to observe the decay of $\eta^\prime\rightarrow\pi^+\pi^-\mu^+\mu^-$.    

\section{BESIII DETECTOR}

The BESIII detector is a magnetic spectrometer~\cite{bes01} located at the Beijing Electron Positron Collider (BEPCII)~\cite{bes02}. The cylindrical core of the BESIII detector consists of a helium-based multilayer drift chamber (MDC), a plastic scintillator time-of-flight system (TOF), and a CsI (Tl) electromagnetic calorimeter (EMC), which are all enclosed in a superconducting solenoidal magnet providing a 1.0 T (0.9 T in 2012, for $1.1\times10^9$ $J/\psi$) magnetic field. The solenoid is supported by an octagonal flux-return yoke with resistive plate counter muon identifier modules interleaved with steel. The acceptance of charged particles and photons is 93\% over 4$\pi$ solid angle. The charged particle momentum resolution at 1 GeV/$c$ is 0.5\%, and the d$E$/dx resolution is 6\% for the electrons from Bhabha scattering. The EMC measures photon energies with a resolution of 2.5\% (5\%) at 1 GeV in the barrel (end cap) region. The time resolution of the TOF barrel part is 68 ps, while that of the end cap part is 110 ps.

\section{DATA SAMPLE AND MONTE CARLO SIMULATION}

The analysis reported here is based on $(1310.6\pm7.0)\times10^{6}$ $J/\psi$ events~\cite{sys01} collected with the BESIII detector in 2009 and 2012.
It is performed in the framework of the BESIII offline software system (BOSS)~\cite{data01} incorporating the detector calibration, event reconstruction and data storage.

The estimation of background and signal efficiency is performed through Monte Carlo (MC) simulations. The BESIII detector is modeled with GEANT4~\cite{data02,data03}. 
The simulation of the production of the $J/\psi$ resonance is performed using the KKMC event generator~\cite{data04,data05}, while the decays are simulated using EVTGEN~\cite{data06,data06-2}. Possible background is studied using a sample of $1.2\times10^9$ simulated $J/\psi$ events in which the known decays of the $J/\psi$ are modelled using the world average values of the branching taken from the Particle Data Group (PDG)~\cite{pdg}, while the unknown decays are generated with the LUNDCHARM model~\cite{data07}.
The final-state radiations (FSR) from charged final-state particles are incorporated with the photos package~\cite{data08}.
An MC simulation with $\eta'\rightarrow\pi^+\pi^-\mu^+\mu^-$ decays uniform over the phase space does not provide a good description of data, therefore a specific generator was developed for this analysis in accordance with the theoretical amplitude in Ref.~\cite{intro09}.

\section{DATA ANALYSIS}
\subsection{Event selection and background analysis}

The final state of interest is studied through the decay chain $J/\psi\rightarrow\gamma\eta', \eta^\prime\rightarrow\pi^+\pi^-\mu^+\mu^-$. Each event is required to contain at least one good photon candidate, and four charged track candidates with a total charge of zero. The MDC provides reconstruction of charged tracks within $|\cos \theta| \leq 0.93$, where the polar angle $\theta$ is defined with respect to the $z$-axis. 
The charge tracks are required to have their point of closest approach to the interaction point (IP) within $\pm 1$ cm in the plane perpendicular to beam direction and within $\pm 10$ cm in beam direction.  
 
Photons are reconstructed from showers in the EMC exceeding a deposited energy of at least 25 MeV in the barrel region ($|\cos \theta| < 0.8$) and 50 MeV in the endcap regions $(0.86 < |\cos \theta|< 0.92)$. The angle between the shower position and the charged tracks extrapolated to the EMC must be greater than 15 degrees. A requirement on the EMC timing is used to suppress electronic noise and energy deposits unrelated to the event.

\begin{figure}[htbp]
  \centering
  \vskip -0.2cm
  \hskip -0.4cm \mbox{
  \begin{overpic}[width=0.24\textwidth]{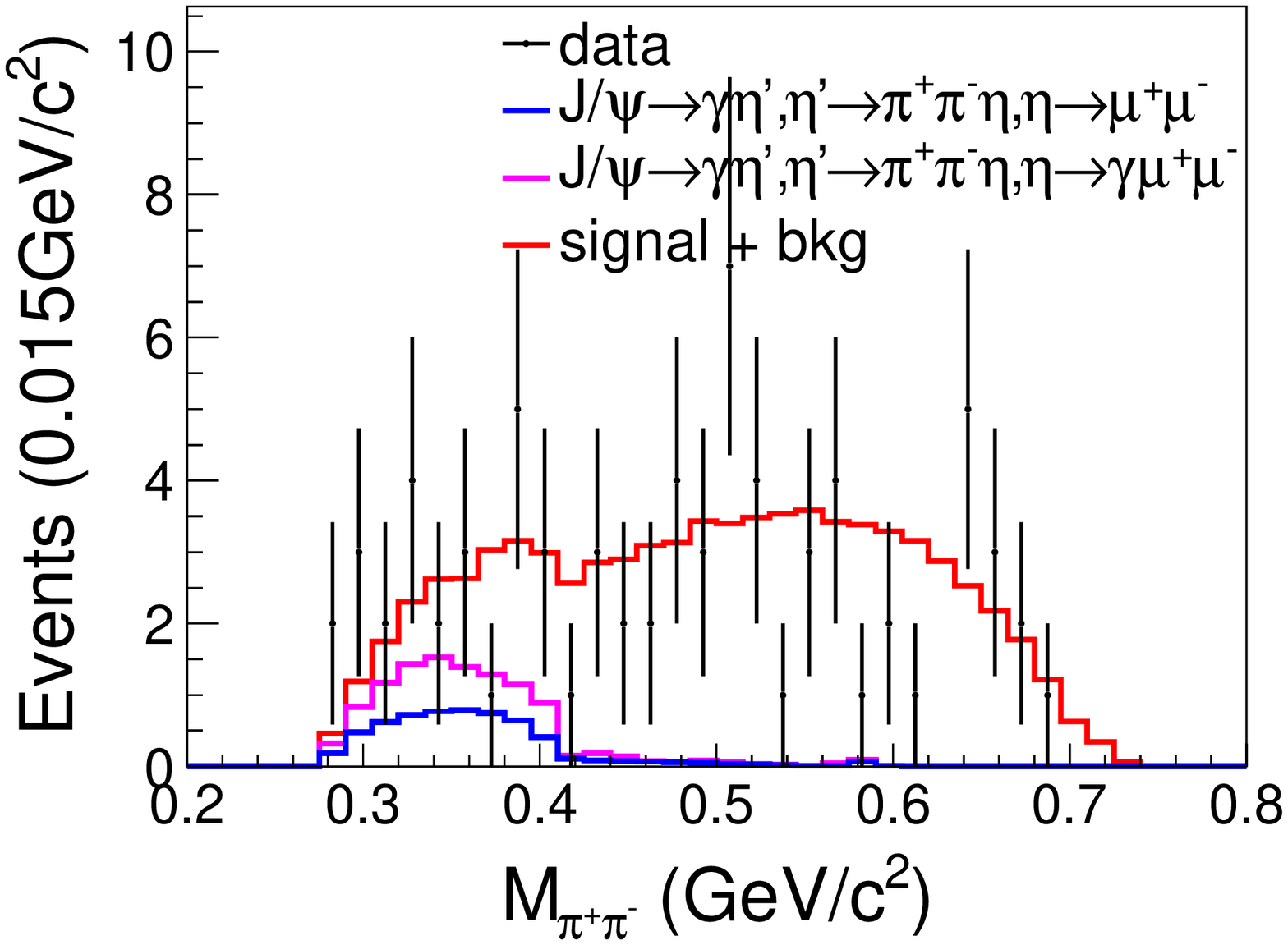}
  \put(20,62){{(a)  }}
  \end{overpic}
  \begin{overpic}[width=0.24\textwidth]{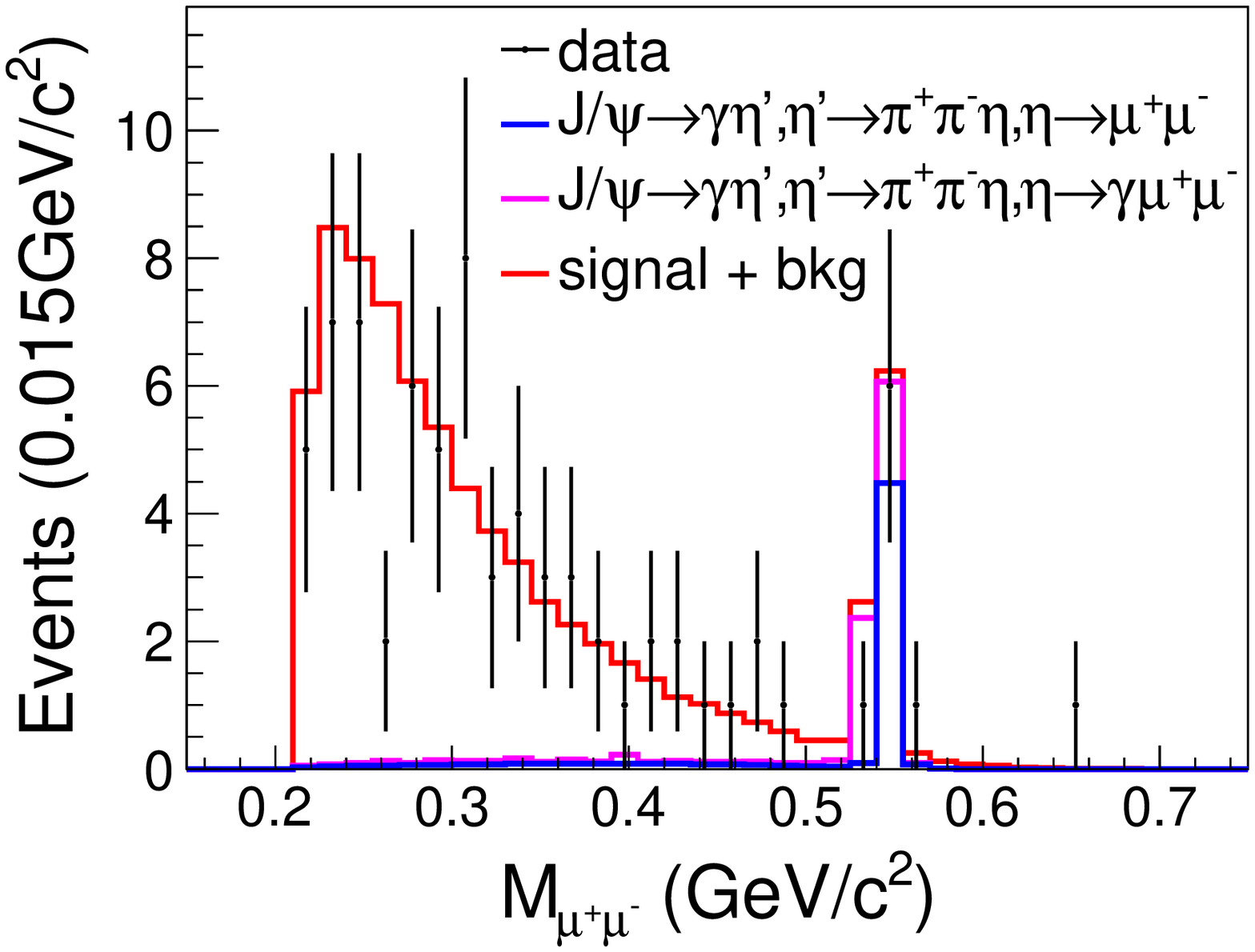}
  \put(20,63){{(b)  }}
  \end{overpic}
  }

  \caption{\label{Mpipi_mumu}
  Invariant mass distribution of (a) $\pi^{+}\pi^{-}$; (b) $\mu^{+}\mu^{-}$ after the event selection. The dots with error bars show data, the red histogram represent signal MC, the pink line is the background $J/\psi\rightarrow\gamma\eta', \eta'\rightarrow\pi^+\pi^-\eta, \eta\rightarrow\gamma\mu^+\mu^-$, and the blue histogram is the background $J/\psi\rightarrow\gamma\eta', \eta'\rightarrow\pi^+\pi^-\eta, \eta\rightarrow\mu^+\mu^-$.
        }
\end{figure}

For each event candidate, TOF and d$E$/dx information are used to perform particle identification (PID) and a four-constraint (4C) kinematic ﬁt imposing energy and momentum conservation is performed under the hypothesis of $\gamma\pi^{+}\pi^{-}\mu^{+}\mu^{-}$.
Here $\chi^2_{4C+\rm PID}=\chi^2_{4C}+\sum_{i=1}^4\chi^2_{{\rm PID}(i)}$ is the sum of the 4C kinematic fit contribution and $\chi^2_{\rm PID}$ produced by combining TOF and and dE/dx information of each charged track for each particle hypothesis (pion, electron, or muon), where i corresponds to the good charged tracks in each hypothesis.
For each event, the hypothesis with the smallest $\chi^2_{4C+{\rm PID}}$ is selected. Events with $\chi^2_{4C}(\gamma\pi^+\pi^-\mu^+\mu^-) < 30$ are kept as $\eta^\prime\rightarrow\pi^+\pi^-\mu^+\mu^-$ candidates. We require that $\chi^2_{4C+{\rm PID}}(\pi^+\pi^-\mu^+\mu^-)$ is less than $\chi^2_{4C+{\rm PID}}(\pi^+\pi^-\pi^+\pi^-)$ to suppress background events from $J/\psi\rightarrow\gamma\pi^+\pi^-\pi^+\pi^-$.  
Possible background events are analyzed with the same procedure using the inclusive MC sample of $1.2\times 10^9$ $J/\psi$ events. The background events mainly originate from the background processes listed in Table~\ref{backgrounds}. For the dominant background channels, the dedicated exclusive MC samples are generated to estimate their contributions to the $\pi^+\pi^-\mu^+\mu^-$ mass spectrum. The corresponding normalized contributions are displayed in Fig.~\ref{fit}.
  \begin {table}[htp]
    \begin{center}
    {\caption {Main background processes and normalized events.}
    \vskip 0.0cm
    \hskip -0.0cm
    \label{backgrounds}}
    \setlength{\tabcolsep}{-1mm}{
    \begin {tabular}{c c}  \hline \hline
      Decay mode       &Normalized events    \\ \hline
   $J/\psi\rightarrow\gamma\eta', \eta'\rightarrow\pi^+\pi^-\eta, \eta\rightarrow\mu^+\mu^-$   &$2$\\
   $J/\psi\rightarrow\gamma\eta', \eta'\rightarrow\pi^+\pi^-\pi^+\pi^-$    &$29$\\
   $J/\psi\rightarrow\gamma\eta', \eta'\rightarrow\pi^+\pi^-\eta, \eta\rightarrow\gamma\mu^+\mu^-$  &$2$\\
   $J/\psi\rightarrow\gamma\eta', \eta'\rightarrow\pi^+\pi^-\eta, \eta\rightarrow\gamma\pi^+\pi^-$ &$2$ \\
   $J/\psi\rightarrow\gamma\eta (1405), \eta (1405)\rightarrow\gamma\phi, \phi\rightarrow\pi^+\pi^-\pi^+\pi^-$ &free\\
   $ J/\psi\rightarrow\gamma\pi^+\pi^-\pi^+\pi^-$  &free\\ \hline \hline
    \end{tabular}}
    \end{center}
    \end{table}
A comparison of the $\pi^+\pi^-$ and $\mu^+\mu^-$ mass spectrum between data and MC in Fig.~\ref{Mpipi_mumu} shows good agreement after requiring 0.94 GeV/$c^2$< $M_{\pi^+\pi^-\mu^+\mu^-}$ < 0.98 GeV/$c^2$. To suppress the background from $\eta\rightarrow\mu^+\mu^-$, we require $|M_{\mu^+\mu^-} - M_{\eta}| > 0.02$ GeV/$c^2$ where $M_\eta$ is the nominal mass of the $\eta$ meson~\cite{pdg}.

A structure corresponding to a $\eta'$ signal is observed in the invariant mass spectrum of $\pi^+\pi^-\mu^+\mu^-$ after applying the above requirements, while the structure around 0.93 GeV/$c^2$ is the background contribution from $J/\psi\rightarrow\gamma\eta', \eta'\rightarrow\pi^+\pi^-\pi^+\pi^-$.

\subsection{Measurement of $\mathcal{B}(\eta^\prime\rightarrow\pi^{+}\pi^{-}\mu^{+}\mu^{-})$}
\begin{figure}[htbp]
  \centering
  \vskip -0.2cm
  \hskip -0.4cm \mbox{
  \begin{overpic}[width=0.42\textwidth]{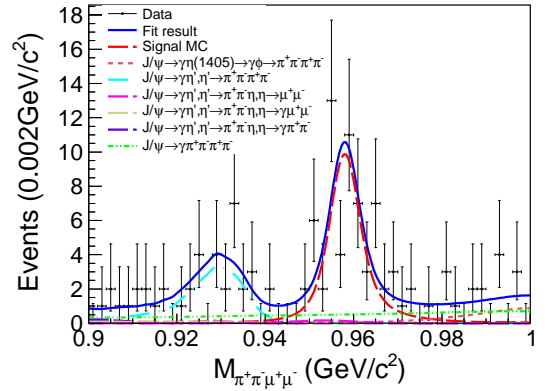}
  %\put(73,80){{(a)  }}
  \end{overpic}
  }
%  \vskip -0.3cm
%  \hskip 0.5cm
  \caption{\label{fit}
   Fit result of the fit to the invariant $\pi^+\pi^-\mu^+\mu^-$ mass. The dots with error bars represent the data, the red line is signal MC and the blue line is the total fit result. The other dotted lines represent background.
  }
\end{figure}

To determine the number of $\eta^\prime\rightarrow\pi^+\pi^-\mu^+\mu^-$ events, an unbinned maximum likelihood fit is performed to the invariant $\pi^+\pi^-\mu^+\mu^-$ mass spectrum. Therefore, the signal shape is determined from signal MC events which are obtained using the DIY generator~\cite{intro09}. For the $\eta'\rightarrow\pi^{+}\pi^{-}\mu^{+}\mu^{-}$ decay, the MC model~\cite{sys04} based on the Vector Meson Dominance Model (VMD model) with finite-width corrections and pseudoscalar meson mixing~\cite{intro07} was developed.
For the backgrounds $\pi^+\pi^-\pi^+\pi^-$ and $\eta(1405)$ the shapes are taken from the MC while the normalization is determined from the fit. The contributions of all other backgrounds are fixed to the MC prediction.
The fit result shown in Fig.~\ref{fit} yields $53\pm9$ signal events.

The statistical significance is determined to be 8$\sigma$. This significance is calculated from the change of the negative log Likelihood function {ln}$\mathcal{L}$ with and without assuming the presence of a signal, while considering the change of degrees of freedom in the fits.

With a detection efficiency of $\epsilon=(39.42\pm0.22)\%$, which is obtained from signal MC simulation, the branching fraction of $\eta^\prime\rightarrow\pi^+\pi^-\mu^+\mu^-$ is calculated as: 
\begin{equation}
\begin{aligned}
\mathcal{B}(\eta'\rightarrow\pi^+\pi^-\mu^+\mu^-)&=\frac{N_{obs}}{N_{J/\psi}\times \mathcal{B}(J/\psi\rightarrow\gamma\eta')\times \epsilon} \\
&= (1.97\pm0.33)\times10^{-5}.
\end{aligned}
\label{f2}
\end{equation}
Here $N_{\rm obs}$ is the signal yield, as determined in the fit, and $\epsilon$ is the detection efficiency for the decay of $\eta^\prime\rightarrow\pi^+\pi^-\mu^+\mu^-$. $\mathcal{B}(J/\psi\rightarrow\gamma\eta')$ is the branching fraction of $J/\psi\rightarrow\gamma\eta'$, $(5.21\pm0.17)\times10^{-3}$\cite{pdg}, and $N_{J/\psi}$ is the number of $J/\psi$ events, $(1310.6\pm7.0)\times10^6$~\cite{sys01}.

\subsection{Systematic uncertainties}

We consider possible sources for systemtatic uncertainty of the branching fraction. Theses systemtatic uncertainties are statistically independent and can be summed up in quadrature, the total systematic uncertainty is $9.3\%$. The corresponding contributions are discussed in detail below and listed in Table~\ref{allsys}.

\begin{itemize}
\item Number of $J/\psi$ events: The number of $J/\psi$ events is determined to be $(1310.6\pm7.0)\times10^6$ from the inclusive hadron events~\cite{sys01}, and the uncertainty of the total number of $J/\psi$ is estimated to be 0.5\%.
\item MDC tracking: The uncertainty due to MDC tracking originates from differences between data and MC. The uncertainty is determined to be 1.0\% per track, using high statistic samples with low background samples of $J/\psi\rightarrow\rho\pi$ and $J/\psi\rightarrow p\bar{p}\pi^+\pi^-$ events~\cite{sys07}. A 4.0\% systematic uncertainty due to MDC tracking efficiency is assigned for the four charged tracks in the decay $\eta'\rightarrow\pi^{+}\pi^{-}\mu^+\mu^-$.
\item Photon detection efficiency: The photon detection efficiency is studied with three independent decay modes, $\psi(2S)\rightarrow\pi^+\pi^-J/\psi (J/\psi\rightarrow\rho^0\pi^0)$, $\psi(2S)\rightarrow\pi^0\pi^0J/\psi (J/\psi\rightarrow l^+l^-)$ and $J/\psi\rightarrow\rho^0\pi^0$~\cite{sys02}. The results indicate that the difference between the detection efficiency of data and MC simulation is within $1.0\%$ for each photon. Therefore, $1.0\%$ is taken to be the systematic uncertainty.
\item PID: The pion PID efficiency for data agrees within $1.0\%$ of that of the MC simulation in the pion momentum region in the analysis~\cite{intro09}. 
There is no specific decay mode available for us to study the PID of muon at low momentum region. MC study shows that the background events of $\eta^\prime\rightarrow \pi^+\pi^-\pi^+\pi^-$ have no contribution to the $\eta^\prime$ peak, which indicates that the pion and muon could be well separated in this specific analysis. Because the mass of the muon is similar to the pion mass, 1.0\% is taken as systematic uncertainty for the muon~\cite{intro09}. Thus, $4.0\%$ is taken as the systematic uncertainty for PID effects.
\item Form factor uncertainty: The MC generator based on the theoretical calculation as explained in Ref.~\cite{sys04} is used to determine the detection efficiency of $\eta'\rightarrow\pi^+\pi^-\mu^+\mu^-$. The detection efficiency dependence on the form factor is evaluated by replacing the form factor above with the form factors introduced in the modified Vector Meson Dominance (VMD) model described in Ref.~\cite{intro07}. The maximum difference of the detection eﬀiciency between hidden gauge model and modified VMD model is determined to be 0.3\% which is taken as the uncertainty due to the form factor, as listed in Table~\ref{allsys}.
\item 4C kinematic fit: The systematic uncertainty from 4C kinematic fit is studied by correcting the track helix parameters to reduce the difference between data and MC simulation~\cite{sys05, sys06}. The detection efficiency from the corrected MC sample is taken as the nominal value, and the difference between the efficiencies with and without correction is determined to be 1.0\% which is taken as systematic uncertainty.
\item Fit range: To estimate the uncertainty from the fit range, we performed alternative fits changing the lower and upper boundaries of the fit range independently by 0.01 GeV/$c^2$. 
Because of the complicated backgrounds at masses larger than 1.0 GeV/$c^2$, the large difference in fit results is obtained for the ranges [0.90-1.01] GeV/$c^2$ and [0.89-1.01] GeV/$c^2$ were obtained. 
The resultant largest difference in the signal yields, 6.2\% is taken as the systematic uncertainty.
\item Background shape: In the fit, the events for three backgrounds ($J/\psi\rightarrow\gamma\eta', \eta'\rightarrow\pi^+\pi^-\pi^+\pi^-$ and $J/\psi\rightarrow\gamma\eta', \eta'\rightarrow\pi^+\pi^-\eta, \eta\rightarrow\gamma\mu^+\mu^-$ and $J/\psi\rightarrow\gamma\eta', \eta'\rightarrow\pi^+\pi^-\eta, \eta\rightarrow\gamma\pi^+\pi^-$) are fixed according to the branching fractions from the PDG~\cite{pdg}. 
To estimate the effect of the uncertainties of the used branching fractions, a set of random numbers has been generated within the uncertainty of each branching fraction. Using these random scaling parameters, a series of fits to the invariant pipimumu mass is performed. The variance of the determined number of signal events is determined to be 1.9\% which is used as systematic uncertainty.
\item Branching fraction of $ J/\psi\rightarrow\gamma\eta'$: The world average branching fraction of $J/\psi\rightarrow\gamma\eta'$, $(5.21\pm0.17)\times10^{-3}$~\cite{pdg}, results in an uncertainty $3.3\%$.
\end{itemize}

\begin{table}[!htbp]
%\begin{small}
  \centering
  \caption{\label{allsys} Sources of systematic uncertainties and their contribution given in \%.}
  %\linespread{1.5}
   % \resizebox{0.50\textwidth}{!}{%
\begin{tabular}{ l c }
  \\
  \hline
  \hline
             Sources                        &~~$\eta'\rightarrow\pi^+\pi^-\mu^+\mu^-(\%)$  \\ \hline
      Number of $J/\psi$ events                                        &~~ $0.5$  \\
      MDC Tracking                                                    &~~ $4.0$  \\
      Photon detection                                                &~~ $1.0$  \\
      PID                                                             &~~ $4.0$  \\
      Form factor uncertainty                                         &~~ $0.3$  \\
      4C kinematic fit                                                &~~ $1.0$  \\
      Fit range                                                       &~~ $6.2$  \\
      Background shape                                                &~~ $1.9$  \\
      $\mathcal{B}(J/\psi\rightarrow\gamma\eta')$                     &~~ $3.3$  \\ \hline
      Total                                                           &~~ $9.3$  \\
  \hline
  \hline
\end{tabular}
 %   }%
%\end{small}
\end{table}

\section{SUMMARY}

With a sample of $1.31 \times 10^9$ $J/\psi$ events, the decay of $\eta^\prime\rightarrow\pi^+\pi^-\mu^+\mu^-$ is observed with a statistical significance of 8$\sigma$ via the process $J/\psi\rightarrow\gamma\eta'$. The branching fraction of $\eta'\rightarrow\pi^+\pi^-\mu^+\mu^-$ is determined to be $\mathcal{B}(\eta'\rightarrow\pi^+\pi^-\mu^+\mu^-)$=(1.97$\pm$0.33(stat.)$\pm$0.18(syst.))$\times10^{-5}$, which is in good agreement with theoretical predictions~\cite{intro05,intro06,intro07}. In addition, the agreement of the generated signal MC with data in Fig.~\ref{Mpipi_mumu} indicates that the theoretical model used, is able to describe the intermediate process resonably. Especially, the expected decreasing spectrum of the dimuon mass in Fig.~\ref{Mpipi_mumu} (b) confirms this further.

\begin{acknowledgments}
The BESIII collaboration thanks the staff of BEPCII and the IHEP computing center for their strong support. This work is supported in part by National Key Basic Research Program of China under Contract No. 2015CB856700; National Natural Science Foundation of China (NSFC) under Contracts Nos. 11625523, 11635010, 11675184, 11735014, 11822506, 11835012, 11935015, 11935016, 11935018; the Chinese Academy of Sciences (CAS) Large-Scale Scientific Facility Program; Joint Large-Scale Scientific Facility Funds of the NSFC and CAS under Contracts Nos. U1632107, U1732263, U1832207; CAS Key Research Program of Frontier Sciences under Contracts Nos. QYZDJ-SSW-SLH003, QYZDJ-SSW-SLH040; 100 Talents Program of CAS; INPAC and Shanghai Key Laboratory for Particle Physics and Cosmology; ERC under Contract No. 758462; German Research Foundation DFG under Contracts Nos. Collaborative Research Center CRC 1044, FOR 2359, GRK 214; Istituto Nazionale di Fisica Nucleare, Italy; Ministry of Development of Turkey under Contract No. DPT2006K-120470; National Science and Technology fund; Olle Engkvist Foundation under Contract No. 200-0605; STFC (United Kingdom); The Knut and Alice Wallenberg Foundation (Sweden) under Contract No. 2016.0157; The Royal Society, UK under Contracts Nos. DH140054, DH160214; The Swedish Research Council; U. S. Department of Energy under Contracts Nos. DE-FG02-05ER41374, DE-SC-0012069

\end{acknowledgments}


\begin{thebibliography}{99}

\bibitem{intro03} E. Witten, Nucl. Phys. B {\bf 223}, 422 (1983).
\bibitem{intro05} A. Faessler, C. Fuchs and M. I. Krivoruchenko, Phys. Rev. C {\bf 61}, 035206 (2000).
\bibitem{intro06} B. Borasoy and R. Nissler, Eur. Phys. J. A {\bf 33}, 95 (2007).
\bibitem{intro07} T. Petri, arXiv:1010.2378 [nucl-th].
\bibitem{intro09} M. Ablikim {\it et al.} (BESIII Collaboration), Phys. Rev. D {\bf 87}, 092011 (2013).
\bibitem{intro08} P. Naik {\it et al.} (CLEO Collaboration), Phys. Rev. Lett. {\bf 102}, 061801 (2009).
\bibitem{sys01} M. Ablikim {\it et al.} (BESIII Collaboration), Chin. Phys. C {\bf 41}, 013001 (2017).
\bibitem{bes01} M. Ablikim {\it et al.} (BESIII Collaboration), Nucl. Instrum. Meth. A {\bf 614}, 345 (2010).
\bibitem{bes02} C. H. Yu {\it et al.}, 10.18429/JACoW-IPAC2016-TUYA01.
\bibitem{data01} W. D. Li, H. M. Liu {\it et al.}, in {\it proceeding of CHEP06, Mumbai, India, 2006,} edited by Sunanda Banerjee (Tata Institute of Fundamental Reserach, Mumbai, 2006).
\bibitem{data02} S. Agostinelli {\it et al.}, Nucl. Instrum. Meth. A {\bf 506}, 250 (2003).
\bibitem{data03} J. Allison, K. Amako, J. Apostolakis, H. Araujo, P. Dubois {\it et al.}, IEEE Trans. Nucl. Sci. {\bf 53}, 270 (2006).
\bibitem{data04} S. Jadach, B. Ward, Z. Was, Comput. Phys. Commun. {\bf 130}, 260 (2000).
\bibitem{data05} S. Jadach, B. Ward, Z. Was, Phys. Rev. D {\bf 63}, 113009 (2001).
\bibitem{data06} R. G. Ping {\it et al.}, Chin. Phys. C {\bf 32}, 599 (2008).
\bibitem{data06-2} D. J. Lange, Nucl. Instrum. Meth. A {\bf 462}, 152 (2001).
\bibitem{pdg} M. Tanabashi {\it et al.} (Particle Data Group), Phys. Rev. D {\bf 98}, 030001 (2018).
\bibitem{data07} J. C. Chen {\it et al.}, Phys. Rev. D {\bf 62}, 1 (2000).
\bibitem{data08} E. Richter-Was, Phys. Lett. B {\bf 303}, 163 (1993).
\bibitem{sys04} Z. Y. Zhang, L. Q. Qin, S. S. Fang, Chin. Phys. C {\bf 36}, 926 (2012).
\bibitem{sys07} M. Ablikim {\it et al.} (BESIII Collaboration), Phys. Rev. D {\bf 85}, 092012 (2012).
\bibitem{sys02} M. Ablikim {\it et al.} (BESIII Collaboration), Phys. Rev. D {\bf 83}, 112005 (2011).
\bibitem{sys05} M.Benayoun, P. David, L. DellBuono, and O.Leitner, Eur. Phys. J. C {\bf 65}, 211 (2010).
%\bibitem{sys05} M.Benayoun, P. David, L. DellBuono, and O.Leitner, Eur. Phys. J. C {\bf 65}, 211 (2010), 0907.4047.
\bibitem{sys06} M. Ablikim {\it et al.} (BESIII Collaboration), Phys. Rev. D {\bf 87}, 012002 (2013).

\end{thebibliography}
\end{document}